\documentclass[apjl]{emulateapj}
\usepackage[below]{placeins}
\usepackage{natbib}
\slugcomment{Submitted to the Astrophysical Journal Letters}

\def\gtorder{\mathrel{\raise.3ex\hbox{$>$}\mkern-14mu
    \lower0.6ex\hbox{$\sim$}}}
\def\ltorder{\mathrel{\raise.3ex\hbox{$<$}\mkern-14mu
    \lower0.6ex\hbox{$\sim$}}}

\shorttitle{Force-feeding Black Holes}
\shortauthors{Begelman}

\begin{document}

\title{Force-feeding Black Holes}

\author{ 
Mitchell C. Begelman\altaffilmark{1} 
}
\affil{JILA, 
     University of Colorado and National Institute of Standards and Technology, \\ 
     440 UCB,  Boulder, CO 80309-0440, USA  \\ 
     email: {\tt mitch@jila.colorado.edu}}

\altaffiltext{1}{Also at Department of Astrophysical and Planetary Sciences, University of Colorado Boulder}

\begin{abstract}
We propose that the growth of supermassive black holes is associated mainly with brief episodes of highly super-Eddington infall of gas (``hyperaccretion").  This gas is not swallowed in real time, but forms an envelope of matter around the black hole that can be swallowed gradually, over a much longer timescale.  However, only a small fraction of the black hole mass can be stored in the envelope at any one time.  We argue that any infalling matter above a few per cent of the hole's mass is ejected as a result of the plunge in opacity at temperatures below a few thousand degrees K, corresponding to the Hayashi track.  The speed of ejection of this matter, compared to the velocity dispersion $\sigma$ of the host galaxy's core, determines whether the ejected matter is lost forever or returns eventually to rejoin the envelope, from which it can be ultimately accreted.  The threshold between matter recycling and permanent loss defines a relationship between the maximum black hole mass and $\sigma$ that resembles the empirical $M_{\rm BH}-\sigma$ relation.  
\end{abstract}

    
\section{Introduction}
\label{sec:intro}

The discovery of a luminous quasar at a redshift of 7.1 (Mortlock et al.~2011) adds urgency to understanding how supermassive black holes grew so large so early.  To grow a $10^9 \ M_\odot$ black hole by Eddington-limited accretion with an efficiency of 0.1
would require continuous growth for 300--650 Myr, depending on whether the initial mass is closer to $10^6$ or 100 $M_\odot$.  At $z=7.1$, this is a substantial fraction of the time available (770 Myr) since the big bang.

A precondition for supermassive black hole growth is the rapid accumulation of gas in the nucleus of the host galaxy.
This is readily accomplished via large-scale, runaway gravitational torques 
(the `bars within bars' instability: Shlosman et al.~1989),
possibly triggered by mergers (Di Matteo et al.~2005; Hopkins et al.~2006; Wise et al.~2008; Levine et al.~2008).
While it is generally agreed that fragmentation can be suppressed when the virial temperature is slightly above the thermal temperature of the gas ($T_{\rm vir} \ga 10^4$ K), it is plausible that a substantial fraction of the infalling matter can reach the vicinity of the black hole even when the gas is able to cool well below $T_{\rm vir}$, due to the transfer of infall energy to turbulent motions (Wise et al.~2008; Begelman \& Shlosman 2009; Regan \& Haehnelt 2009; Mayer et al.~2010).  

These mass supply mechanisms do not ``know" about the Eddington limit of the black hole.  The characteristic mean inflow rate associated with self-gravitating infall, 
\begin{equation}
\label{mdot}
\dot M_{\rm in} \sim \sigma^3/ G = 2000 \sigma_{200}^3 \ M_\odot \ {\rm yr}^{-1}, 
\end{equation}
where $200 \sigma_{200}$ km s$^{-1}$ is the velocity dispersion in the central regions of the galaxy, typically exceeds the Eddington accretion rate of the black hole by $\gtorder 2-3$ orders of magnitude.  Black holes in this situation might be considered to be `force fed'.

What happens to this gas as it nears the black hole? The infalling gas is likely to be highly dissipative and radiative.  At radii below 
\begin{equation}
\label{rtrap}
R_{\rm tr} \sim {\kappa \dot M_{\rm in}\over 4\pi c} \sim 10^{17} \sigma_{200}^3 \ {\rm cm} 
\end{equation}
(where the opacity, $\kappa$, is assumed to be comparable to the electron scattering opacity for fully ionized gas,  $0.34$ cm$^2$ g$^{-1}$) the gas traps its own radiation (Begelman 1978) and radiation pressure forces become comparable to gravity and dynamical stresses. (This `trapping radius' will fluctuate if the inflow rate is unsteady.) The pressurized inflow becomes less responsive to gravitational torques (and gravitational instability) but possibly more capable of transferring angular momentum through internal stresses, e.g., due to magnetic fields or turbulence.  

Once inside the trapping radius, the flow radiates inefficiently.  The outward energy flux associated with angular momentum transport causes the gas to become unbound (Narayan \& Yi 1994, 1995), unless the inflow rate is suppressed.  This may trigger winds that reduce the inflow rate at every radius to the point where the radiation is barely trapped (Shakura \& Sunyaev 1973; Blandford \& Begelman 1999, 2004).  Hyperaccreting sources such as the Galactic binary SS433 (King et al.~2000; Begelman et al.~2006b), Galactic microquasars during outburst (Mirabel \& Rodriguez 1999), and possibly ultraluminous X-ray sources (King 2008; Gladstone et al.~2009) all appear to produce powerful outflows consistent with this outcome.    

Powerful mass loss does not favor black hole growth since the gas is being supplied at a much higher rate than it can be absorbed.  If the excess mass is permanently lost, then the increase of black hole mass per infall episode is given by the Eddington-limited accretion rate times the duration of the infall.  The total baryonic matter contained within the virial radius of a galactic bulge is about $M_b \sim 2 \times 10^{10} \sigma_{200}^4\ M_\odot$ (assuming a baryon fraction of $\sim 0.2$), and if we assume that as much as 10 per cent of this matter reaches the black hole, then the effective duration of a hyperaccretion episode (which is likely to be intermittent, with a fairly small duty cycle) is $t_{\rm in} \sim 0.1 M_b/\dot M_{\rm in} \sim 10^6 \sigma_{200}$ yr.  This is inadequate to double the black hole's mass, even if 
the black hole accretes at $\sim 6$ times the usual Eddington rate,
which is plausible since the accretion rate can exceed the Eddington rate by a factor $\sim \ln (1 + \dot M_{\rm in}/\dot M_E) $ if the flow is able to radiate through a rotational funnel (King 2008).  In contrast, the amount of available gas would be enough to increase the black hole mass twenty-fold, if most of it could be retained and ultimately accreted.  Begelman (2012) has argued that such retention is a possible outcome of steady-state hyperaccretion. 

In this Letter, we discuss the radiation hydrodynamics of force-fed envelopes around black holes. In section 2 we outline the basic properties of such envelopes, assuming that the opacity is roughly constant.  This assumption is relaxed in section 3, where we argue that the sharp drop in opacity below $\sim 3000$ K prevents an envelope from remaining in dynamical equilibrium if it exceeds a certain radius and mass.  Attempting to add mass beyond this limit leads to such vigorous mass ejection that the mass and radius should be regulated at close to the limiting value.  One implication of this self-regulation is explored in section 4, where we suggest that the question of whether mass is permanently or temporarily expelled from the envelope may provide an explanation for the $M_{\rm BH}-\sigma$ relation linking black hole mass to galaxy velocity dispersion.  In section 5, we discuss the results and consider the observational appearance of force-fed black holes.   

\section{Force-fed envelopes}
\label{sec:ff}

Consider a radiation pressure-supported envelope of mass $M_*$ surrounding a black hole of mass $M \gg M_*$. 
Whether the envelope is built up gradually by steady inflow at rate $\dot M_{\rm in}$ or intermittently by the arrival of discrete clumps of gas,
the system radiates at roughly the Eddington limit $L_E = 4\pi GM c/\kappa$ for the black hole.  Suppose that the gas rearranges its internal angular momentum distribution on a few dynamical timescales, so that substantial portions of the matter begin to move inward.  Gravitational binding energy liberated in excess of $L_E$ (modulo geometric factors of order a few) is unable to escape and increases the internal energy of the gas, which is then pushed outward by pressure forces.  The spreading of the gas outwards tends to correct for the overproduction of energy.  If this condition applies at every radius, including the inner region where gas liberates energy as it accretes on to the black hole, then the overall structure of the envelope will regulate itself so that energy production is approximately in balance with global energy loss.  Assuming that newly infalling matter is captured at a radius smaller than that of the envelope, we can neglect rotational support on large scales and treat the envelope as roughly spherical. 

Because the black hole is accreting at roughly the Eddington limit, the radiative diffusion time scale, $t_{\rm diff} \sim \rho \kappa R^2/c$, is comparable to $t_{\rm dyn}$ at a few Schwarzschild radii. Suppose the density varies $\propto R^{-n}$ with increasing distance from the hole.  If $n>1/2$, the ratio $t_{\rm diff}/ t_{\rm dyn}$ decreases with $R$ and the radiation would leak out of the system on too short a time scale for it to be replenished.  This situation would presumably lead to a thin disk accreting very slowly.  On the other hand, if $n<1/2$ the radiation would remain trapped but the system would become violently unstable to convection. To see this, note that $p\propto \rho/R$ in hydrostatic equilibrium, hence the entropy function $p/\rho^{4/3} \propto R^{n/3 - 1}$ decreases with radius.  The resulting convection would be saturated, with a convective luminosity $L_c \sim p (p/\rho)^{1/2} R^2 \propto R^{1/2 - n}$ that increases with $R$.  In this case, the rapid loss of energy would steepen the density profile until it approached $R^{-1/2}$. This marginal case, with $t_{\rm diff}/ t_{\rm dyn} \sim O(1)$ at all $R$, appears to be the only plausible scaling for a low-mass envelope supported by accretion luminosity.   Analogous arguments give the same radial scalings for density and pressure in various manifestations of radiatively inefficient accretion flows, whether the main energy transport mechanism is convection (Narayan et al.~2000; Quataert \& Gruzinov 2000), organized outflows (Blandford \& Begelman 1999), or saturated thermal conduction (Gruzinov 1998).  The main difference here is that the flow is radiation pressure-dominated and adjusts to radiate away the dissipated energy.  These arguments are independent of whether the angular momentum distribution is quasi-Keplerian (as indeed it is not in the Gruzinov 1998 version, which assumes non-rotating, spherically symmetric accretion).  The scalings in the outer envelope are also unaffected by the efficiency with which energy is released in the inner zone. This observation applies particularly to the case where the final plunge into the black hole occurs from orbits of low binding energy (close to the marginally bound circular orbit: e.g., Abramowicz et al.~2010).  In this case, the density and pressure profile steepen in the central regions to allow a larger net accretion rate, but the convective luminosity is still regulated by the condition of global radiative equilibrium. 

We set $L_c = 4\pi \beta R^2 p^{3/2} \rho^{-1/2} = \eta L_E$, where $\beta \la 1$ is a convective efficiency parameter.  If the envelope radiated at exactly the Eddington limit for the total mass, then $\eta \approx M_*/M \ll 1$ throughout most of the interior.  More realistically, we expect the total energy flux to be somewhat larger than $L_E$, due to the strong density fluctuations that are likely to be associated with convection (resulting in a ``porous" atmosphere: Shaviv 1998) as well as possible super-Eddington losses (which can increase with the logarithm of the radius) through a rotational funnel (Jaroszy\'nski et al.~1980; Paczy\'nski \& Wiita 1980; Sikora 1981).  We therefore consider $\eta \ga 1$.  
The density is 
\begin{equation}
\label{rhor}
\rho (R) = \left( {3\over 2}\right)^{3/2} {\eta \over \beta}{c\over \kappa (GMR)^{1/2} }, 
\end{equation}
with the pressure given by $p = (2/3) GM \rho(R)/R$.
Integrating the density and setting the mass equal to $M_*$, we obtain the radius of the envelope:
\begin{equation}
\label{Rstar}
R_* = 3.8 \times 10^{18} \left( {\eta\over \beta}\right)^{2/5} \left( {\kappa\over \kappa_{\rm es}}\right)^{2/5} \left({M_*\over M}\right)^{2/5} M_8^{3/5} \ {\rm cm},
\end{equation}
where $M_8 = M/10^8 \ M_\odot$ and $\kappa_{\rm es}$ is the electron scattering opacity.

\section{Maximum mass and radius}
\label{sec:maxmass}

Consider a force-fed envelope with a radius given by eq.~(\ref{Rstar}), radiating at $\eta_{\rm rad}L_E$.  The effective temperature of the photosphere (assumed to be located close to $R_*$) is then
\begin{equation}
\label{Teff}
T_{\rm eff} = 1.1 \times 10^{3} \eta_{\rm rad}^{1/4} \left({\beta\over \eta}\right)^{1/5} \left({M_*\over M}\right)^{-1/5} M_8^{-1/20} \ {\rm K},
\end{equation}
where we have taken $\kappa \approx \kappa_{\rm es}$.  For a given $M$, $T_{\rm eff}$ is a decreasing function of envelope mass.  If $T_{\rm eff}$ were to fall below some value $T_{\rm min}$ between about 2000 K and 4500 K, depending on metallicity, the opacity would drop so sharply with decreasing temperature (assuming that the photospheric radiation is close to LTE) that no photospheric solution could be found. $T_{\rm min}$, of course, represents the Hayashi track that determines the minimum temperatures of red giants and contracting protostars.  In the case of force-fed envelopes, reaching the Hayashi track apparently leads to the loss of dynamical equilibriium.

To see why, imagine a static envelope that violates the condition $T_{\rm eff} > T_{\rm min}$.  Defining $T_3 \equiv T_{\rm min}/ 3000$ K, the condition is violated for 
\begin{equation}
\label{Mmax}
{M_* \over M} > 6.6 \times 10^{-3}\eta_{\rm rad}^{5/4}\left( {\beta\over \eta}\right)T_3^{-5} M_8^{-1/4} .
\end{equation}  
Since we expect $\beta / \eta < 1$, the maximum envelope mass $M_H$ consistent with the Hayashi track is smaller than $M$ for all $T_{\rm min} > 2000 K$ and $M_8 \gtorder 0.01$, even if $\eta_{\rm rad}$ is somewhat larger than one.
The condition $T_{\rm eff} > T_{\rm min}$ cannot actually be violated, because any matter with $T<T_{\rm min}$ has such a low opacity that it cannot be supported against gravity by the radiation flux emerging from the envelope.  What must happen, instead, is that the radius of the envelope no longer increases with envelope mass for $M_* > M_H$, but rather levels off at a value $R_H \sim (L /4\pi \sigma_{\rm SB}T_{\rm min}^4)^{1/2}$, where $L$ is the radiative flux crossing the photosphere.

As we have seen, increasing the envelope mass at fixed radius increases the outward convective transport of energy. It probably also increases the rate at which the envelope liberates gravitational binding energy.  In particular, the rate at which matter is funneled into the black hole is likely to increase.  When $M_* > M_H$, then, the convectively transported luninosity is expected to exceed $L_E$ by roughly a factor $M_*/M_H$.  When the opacity is constant, this excess luminosity leads to excess pressure that inflates the envelope, bringing the luminosity back down to $\sim L_E$.  But this pressure release valve is not available at effective temperatures below $T_{\rm min}$.  Thus, the envelope continues to produce a luminosity in excess of $L_E$. 

Escaping luminosity substantially in excess of $L_E$ must drive a wind, even if the gas in the outermost layers is close to $T_{\rm min}$ and therefore has a low opacity.  This is because the transition from convective radiation transport to radiative diffusion does not occur at the photosphere, but rather at an optical depth $\tau_c \sim c/v_c \gg 1$, where the temperature (in the limit of a static, plane-parallel atmosphere) is $\sim \tau_c^{1/4} T_{\rm min}$.  Since the opacity is such a steep function of $T$ near $T_{\rm min}$, the Eddington limit at the base of the radiative zone should be estimated using the electron scattering opacity.

Envelope mass in excess of $M_H$ becomes unbound on a few dynamical timescales.  For a limiting photospheric radius
\begin{equation}
\label{Rmax}
R_H \sim 5 \times 10^{17} T_3^{-2} M_8^{1/2} \ {\rm cm} ,
\end{equation} 
the characteristic mass loss rate is $\sim M_H (GM/R_H^3)^{1/2} \sim 10^4 T_3^{-2} M_8^{1/2}M_\odot$ yr$^{-1}$, considerably higher than typical values of $\dot M_{\rm in}$.   This implies strong feedback that keeps the envelope mass from exceeding $M_H$ by a large factor.  The corrective mass loss is probably not continuous for envelope masses slightly above $M_H$, since this would imply a luminosity only slightly higher than $L_E$.  Since the radiation pressure force effectively cuts off close to the photosphere, due to the decrease in opacity, it requires a luminosity of a few times $L_E$ to drive mass loss at a high enough speed to escape the gravitational potential of the black hole.  This suggests that the feedback is episodic, with envelope masses growing to perhaps a few $M_H$ before suffering a major mass loss event.

\section{The $M_{\rm BH}-\sigma$ relation}
\label{sec:Msigma}

The self-regulatory behavior of force-fed envelopes suggests a new explanation for the relationship between the final black hole mass and $\sigma$ (Ferrarese \& Merritt 2000; Gebhardt et al.~2000). Matter expelled through the photosphere must have at least the escape speed, $(2 GM/R_{\rm H})^{1/2}$, to escape the potential of the black hole, since this matter experiences little additional radiative driving farther out.  This is probably guaranteed by the launching process.  However, the outflow cannot know about the galactic potential that becomes dominant beyond the Bondi radius, $R_B = GM/\sigma^2$.  To escape from this potential, gas has to reach the Bondi radius with a speed $v_B$ that satisfies 
\begin{equation}
\label{vB}
v_B^2 > 4\sigma^2 \left[ 1 +  \ln \left( {R_o \over R_B } \right) \right] ,
\end{equation} 
where we have assumed a singular isothermal potential with a one-dimensional velocity dispersion $\sigma$, out to a radius of $R_o$, and a Keplerian potential beyond.  Let us suppose that the wind is launched from a radius $\xi R_H$ with a speed $v_W = (\chi GM/ \xi R_H)^{1/2}$.  Since $R_H$ is the radius of a photosphere radiating a luminosity $L_E$ with effective temperature $T_{\rm min}$, and we have argued that $L$ must exceed $L_E$ by a factor of a few in order to drive the wind, we expect $\xi > 1 $.  The condition for escape of the wind from the galaxy is then   
\begin{equation}
\label{vB2}
{GM\over R_H }> {4\xi\over \chi - 2 } \sigma^2 \left[ 1 +  \ln \left( {R_o \over R_B } \right) \right] .
\end{equation}        

Anticipating that the black hole mass is about 0.1 per cent that of the galactic core, we take $R_o/R_B \sim 10^3$ within the logarithmic factor.  Substituting for $R_H$ from eq.~(\ref{Rmax}), we obtain the escape condition 
\begin{equation}
\label{Msigma}
M > M_{\rm esc} = 2\times 10^7 \left({\xi\over \chi - 2 }\right)^2 T_3^{-4}\sigma_{200}^4  \ M_\odot.
\end{equation}        
We interpret this inequality as follows.  If $M> M_{\rm esc}$, then any gas added to a force-fed envelope in excess of $M_H$ is flung out of the galaxy, never revisits the vicinity of the black hole, and will never be accreted.  On the other hand, if $M < M_{\rm esc}$, any matter expelled from the envelope is nevertheless trapped in the galactic potential and, provided it does not condense into stars, is available to be accreted in the future. 

If the matter needed for black hole growth is supplied during infrequent episodes of rapid infall, it may be difficult to grow black holes to a mass exceeding $M_{\rm esc}$.  As long as the black hole mass is less than $M_{\rm esc}$, essentially all infalling matter (modulo matter lost to star formation) is ultimately available for accretion.  Matter expelled from the force-fed envelope keeps returning until it is accreted, presumably at a self-regulated rate (which may exceed the usual Eddington limit if the accreted gas has a low binding energy or if substantial energy escapes through a porous atmosphere or polar jet).  But this is not the case for black holes whose masses exceed $M_{\rm esc}$.   Once the infall stops, the black hole drains the existing envelope but does not receive a large supply of fuel thereafter, until the next episode of rapid infall begins.  Thus the duty cycle of accretion drops sharply for black holes once their masses exceed $M_{\rm esc}$ and we may regard this as an upper mass limit. 

For plausible values $\xi^2 (\chi-2)^{-2} T_3^{-4} \sim$ a few, expression (10) for $M_{\rm esc}$ closely matches the empirical $M_{\rm BH} - \sigma$ relation, in both slope and normalization (Tremaine et al.~2002).

\section{Discussion and conclusions}
\label{sec:disc}

We have discussed the accumulation of force-fed envelopes around black holes being supplied with matter at a highly super-Eddington rate.  The existence of such envelopes is predicated upon two conditions: first, that the feedback from energy liberated close to the black holes is able to keep the envelope inflated; and second, that inflowing gas releasing energy in excess of $L_E$ is not immediately ejected. The former condition requires that radiation remain at least partially trapped in the gas as it accumulates, which places a lower limit on the density as a function of radius. Begelman (2012) has demonstrated the possibility of quasi-Keplerian ``inflow-outflow" (ADIOS) models that satisfy the second condition. Here we have focused on envelopes that have inflated to the extent that rotation is dynamically unimportant, and have argued that the convective properties of such flows select a special set of solutions where the density scales as $\rho \propto R^{-1/2}$.   

Force-fed envelopes, which have masses $M_* \la M$, should be distinguished from ``quasi-stars" (Begelman et al.~2006a, 2008), which are  accretion-powered convective envelopes of mass $M_* \gg M$ surrounding recently formed black holes.  The convection in quasi-stars is unsaturated, allowing them to be modeled as $n=3$ ($\gamma = 4/3$) polytropic envelopes. Ball et al.~(2011, 2012) have shown that quasi-stars have a {\it lower} mass limit, $M_* \ga 10 M$, due to the impossibility of matching a finite polytropic envelope to the black hole-dominated core. Opacity effects, similar to those discussed above, may also limit quasi-star masses to $\ga 10-100 M$ (Begelman et al.~2008).  Estimates relying on hypothetical wind mass loss rates (Dotan et al.~2011) yield a similar lower limit. 

Force-fed envelopes exhibit an {\it upper} mass limit, driven by the temperature-dependence of opacity.  The reason opacity gives an upper mass limit here, as opposed to a lower limit for quasi-stars, is that the convection in force-fed envelopes is saturated.  For typical conditions, this limit is a fraction of the black hole mass.  Thus, there is a gap in allowed envelope masses: it appears impossible to re-create a quasi-star by accretion, once it has dispersed.  It is also impossible for a force-fed envelope to store enough matter to grow a large black hole rapidly.  If such envelopes are implicated in the rapid growth of supermassive black holes, it must be because they disperse through relatively slow winds, which can be captured by the surrounding galaxy's potential and recycled until the mass is ultimately accreted.  We have shown that a simple recycling condition implies a maximum black hole mass that approximately reproduces the empirical $M_{\rm BH} - \sigma$ correlation.  This proposed explanation ties the $M_{\rm BH} - \sigma$ relation to atomic physics through the same mechanism that sets the minimum temperature (Hayashi track) of red giant photospheres and contracting protostars.  

During their enveloped phase, force-fed black holes would not resemble normal AGN or even highly obscured (Compton-thick) AGN.  The optical depth through the envelope, $\tau \sim c(R_H/GM)^{1/2}$, would imply column densities as large as hundreds of g cm$^{-2}$, enough to degrade hard X-rays from the core.  Their photospheres would have temperatures similar to red giants.  On the other hand, the inevitable rotational funnels close to the black hole might well produce fast jets that penetrate the envelope and emit hard radiation.     

\acknowledgements
I thank Phil Armitage for helpful discussions. This work was supported in part by NSF grant AST-0907872.

\end{document}